\documentclass[aps,showpacs,twocolumn,
superscriptaddress]{revtex4}
\usepackage{graphicx}% Include figure files
\usepackage{dcolumn}% Align table columns on decimal point
\usepackage{bm}% bold math
\usepackage{color}
\usepackage[normalem]{ulem} % \sout{old text} for strikeout
\usepackage[dvipsnames]{xcolor} % For blue in-text comments
\usepackage{hyperref}
\hypersetup{
  colorlinks=true,        % false: boxed links; true: colored links
  linkcolor=blue,         % color of internal links
  citecolor=cyan,         % color of links to bibliography
}
\usepackage{mathrsfs}
\usepackage[utf8]{inputenc}
\usepackage{mathtools}
\usepackage{doi}
\usepackage{amsmath}
\usepackage{amssymb}

\begin{document}

\title{Dynamics of charged and magnetized particles around \\ cylindrical black holes immersed in external magnetic field}

\author{Rayimbaev Javlon}
\email{javlon@astrin.uz}
\affiliation{Ulugh Beg Astronomical Institute, Astronomicheskaya 33, Tashkent 100052, Uzbekistan}
\affiliation{National University of Uzbekistan, Tashkent 100174, Uzbekistan}
\affiliation{Institute of Nuclear Physics, Ulugbek 1, Tashkent 100214, Uzbekistan}
\author{Demyanova Alexandra}
\email{demyanova@astrin.uz}
\affiliation{Ulugh Beg Astronomical Institute, Astronomicheskaya 33, Tashkent 100052, Uzbekistan}

\author{Ugur Camci}
\email{ucamci@rwu.edu}
\affiliation{Department of Chemistry and Physics, Roger Williams University, One Old Ferry Road, Bristol, RI 02809, USA}
\author{Ahmadjon~Abdujabbarov}
\email{ahmadjon@astrin.uz}

\affiliation{Ulugh Beg Astronomical Institute, Astronomicheskaya 33, Tashkent 100052, Uzbekistan}
\affiliation{National University of Uzbekistan, Tashkent 100174, Uzbekistan}
\affiliation{Institute of Nuclear Physics, Ulugbek 1, Tashkent 100214, Uzbekistan}
\affiliation{Shanghai Astronomical Observatory, 80 Nandan Road, Shanghai 200030, P. R. China}
\affiliation{Tashkent Institute of Irrigation and Agricultural Mechanization Engineers, Kori Niyoziy, 39, Tashkent 100000, Uzbekistan}
\author{Bobomurat~Ahmedov}
\email{ahmedov@astrin.uz}
\affiliation{Ulugh Beg Astronomical Institute, Astronomicheskaya 33, Tashkent 100052, Uzbekistan}
\affiliation{National University of Uzbekistan, Tashkent 100174, Uzbekistan}
\affiliation{Tashkent Institute of Irrigation and Agricultural Mechanization Engineers, Kori Niyoziy, 39, Tashkent 100000, Uzbekistan}
\date{\today}

\begin{abstract}
This paper investigates the motion and acceleration of a particle that is electrically and magnetically charged, which rotates around a cylindrical black hole in the presence of an external asymptotically uniform magnetic field parallel to the $z$ axis. Basically, the work considers the circular orbits of particles rotating around the central object and studies the dependence of the most internal stable circular orbits (ISCO) on the so-called magnetic coupling parameters, which are responsible for the interaction between the external magnetic field and magnetized and charged particles. It is also shown that the ISCO radius decreases with increasing magnetized parameter. Therefore, collisions of magnetized particles around a cylindrical black hole immersed in an external magnetic field were also studied, and it was shown that the magnetic field can act as a particle accelerator around non-rotating cylindrical black holes.

\end{abstract}

\maketitle

\section{Introduction}

High-energy radiation processes and energy release near the rotating horizon of a black hole have received more attention in recent publications ~\cite {Dadhich18}. For example, the classical Penrose effect ~\cite {Penrose73} is one such process that has been extensively studied in the literature, for example. in the links ~\cite{Piran75, Piran77, Piran77a}. Moreover, Banados and his co-authors in [~\cite {Banados09}] demonstrated that for an extremely spinning black hole head, collisions can generate high-energy particles of the center of mass (the so-called BSW process). The results of \cite{Banados09} were commented on in \cite{Berti03}, where the authors concluded that astrophysical constraints on maximum spin, back reaction effects, and sensitivity to initial conditions place severe constraints on the likelihood of such acceleration. Energy release and particle acceleration around a rotating black hole in Ho \v{r} Ava-Lifshitz gravity were studied in our previous article in ~\cite{Abdujabbarov11b}. Particle acceleration, circular geodesic, accretion disk, and high-energy collisions in Janice-Newman-Vinicourt spacetime were investigated ~\cite {Patil2012PhRvD, Chowdhury2012PhRvD}. It was recently shown ~ in ~\cite {Stuchlik2012CQG} that Kerr primordial super-spinars, extremely compact objects with an appearance described by the geometry of the bare Kerr singularity, can serve as efficient accelerators for extremely high-energy collisions.

In Ref.~\cite{Frolov12} author has shown that in the existence of the magnetic field innermost stable circular orbits (ISCO) of charged particles can be situated near to the horizon in the ground of Schwarzschild spacetime. i.e shifts towards to center black hole. He also shown that for a collision of two particles, one of which is charged and rotating at ISCO and the other is neutral and falling from infinity, the maximal collision energy formally can be arbitrary high. This result has some affinity with the recently discussed effect of high center-of-mass energy for collision of particle in the vicinity of extremely rotating black holes.

{No-hair theorem tells that no black hole can have its own magnetic field, but one can take into account the assumption that a black hole immersed in an external magnetic field generated by an electric current from the charged matter of an accretion disk or a magnetized object will accompany the black hole.After Wald's solution of the electromagnetic field equation around black holes trapped in an external asymptotically uniform magnetic field~\cite {Wald74}, the different properties of electromagnetic fields around various black holes in external / intrinsic asymptotically uniform magnetic fields and magnetized neutron stars with an appropriate dipolar magnetic field were studied by several authors conditions~\cite{Aliev86,Aliev89,Aliev02,Frolov11,Benavides-Gallego18,Shaymatov18,Stuchlik14a,Abdujabbarov10,Abdujabbarov11a,Abdujabbarov11,Abdujabbarov08,Karas12a,Stuchlik16,Kovar10,Kovar14,Kolos17,Pulat2020PhRvDMOG,Rayimbaev2019IJMPCS,Rayimbaev2020MPLA,Rayimbaev2019IJMPD}.
The dynamics of dipolar magnetized, electrically and magnetically charged particles around various black holes immersed in an external magnetic field, as well as electrically and magnetically charged black holes have already been widely studied in various gravitational theories~\cite{deFelice,deFelice2004,Rayimbaev16,Oteev16,Toshmatov15d,Abdujabbarov14,Rahimov11a,Rahimov11,Haydarov20,Haydarov2020EPJC,Abdujabbarov2020PDU,Narzilloev2020PhRvDstringy,Rayimbaev2020PhRvD,TurimovPhysRevD2020,DeLaurentis2018PhRvD,MorozovaV2014PhRvD,Nathanail2017MNRAS,Vrba2020PhRvD,Vrba2019EPJC}.}

The aim of this paper is to show that a similar effect of particle
collision with high center-of-mass energy is also possible when a
black hole is non-rotating (or slowly rotating) provided there
exists magnetic field in its exterior.  This study might be
interesting since there exist both theoretical \cite{Piotrovich11} and experimental \cite{Baczko2016AnA} indications that such a magnetic field must be present in the vicinity of black holes. In what follows we assume that this field is test one and its energy-momentum does not modify the background
black hole geometry. For a black hole of mass $M$ this condition
holds if the strength of magnetic field satisfies to the
condition~\cite{Frolov12,Piotrovich11}
\begin{equation}
\label{BBB}
B\ll B_{max}={c^4\over G^{3/2}
M_{\odot}}\left(\frac{M_{\odot}}{M}\right)\sim
10^{19}\left(\frac{M_{\odot}}{M}\right)\mbox{Gauss}\, .
\end{equation}
Black holes with such field characteristics are called "weakly magnetized". The condition (\ref {BBB}) can be expected to hold for both stellar mass and supermassive black holes.

The work is structured as follows. Section ~II is devoted to the study of the motion of charged particles in the space-time of a magnetized cylindrical black hole, with the main focus on the properties of the ISCO radii. The mechanism of particle acceleration in the region surrounding a magnetized black hole in cylindrical coordinates is discussed in Section ~III. Concluding remarks and discussions are presented in section ~IV.

Throughout this article, we use a space-time signature like $(-, +, +, +) $ and a system of units in which $ G = 1 = c $ (However, for these expressions with an astrophysical application, we wrote the speed of light explicitly.). Greek indexes are accepted from 0 to 3.

\section{\label{sec:emfield}
Magnetized Black Hole in Cylindrical Coordinates}

It is interesting to study gravitational objects with the cylindrical symmetry from astrophysical point of view, since the cylindrical symmetry can be applied to study the jets and cosmic strings. From theoretical point of view cylindrical symmetry is important to study conical singularities and spacetime defects. 

The rotating charged cylindrical black hole vacuum solution is derived by considering the Einstein-Hilbert action with a cosmological constant in four dimensions~\cite{Lemos1995CQG,Lemos1995PhLB,Lemos1996PhRvD} and inserting a cylindrical symmetric generic metric into this action. The explicit spacetime metric  in the cylindrical coordinates  $(t, r, \phi, z)$ is
\begin{eqnarray}\label{metric1}
ds^2&=&-\Delta(\gamma dt - \frac{\omega}{\alpha^2 }d\phi)^2 +
r^2(\omega dt - \gamma d\phi)^2\nonumber\\&& +\frac{dr^2}{\Delta}
+ \alpha^2 r^2 dz^2\ ,\ \quad \label{metric}
\end{eqnarray}
where metric coefficients take form
\begin{eqnarray}
\Delta&=&\alpha^2r^2-\frac{\beta}{\alpha
r}+\frac{c^2}{\alpha^2r^2}\ ,\nonumber\\
\beta&=&4M\left(1-\frac{3}{2}a^2\alpha^2\right)\ ,\nonumber\\
c^2&=&4Q^2\left(\frac{1-\frac{3}{2}a^2\alpha^2}{1
-\frac{1}{2}a^2\alpha^2}\right)\ ,\nonumber\\
\gamma&=&\sqrt{\frac{1-\frac{1}{2}a^2\alpha^2}{1
-\frac{3}{2}a^2\alpha^2}}\ ,\nonumber\\
\omega&=&\frac{a\alpha^2}{\sqrt{1-\frac{3}{2}a^2 \alpha^2}}\
,\nonumber
\end{eqnarray}
with $\alpha^2=-\frac{1}{3}\Lambda$ results in a real $\alpha$, $\lambda$ is cosmological constant.

Hereafter we will consider the nonrotating black hole (i.e. $a=0$,
so that $\omega=0$) with zero electric charge $Q$ immersed in
external asymptotically uniform magnetic field. Then we may
rewrite the metric (\ref{metric}) in the following form
\begin{eqnarray}\label{metric1}
\label{metric1} ds^2&=&-\Delta dt^2 +\frac{dr^{2}}{\Delta} + r^2
d\phi^2 + \alpha^2 r^2 dz^2\ ,
\end{eqnarray}
where $\Delta=\alpha^2r^2-{\beta}/{(\alpha r)}$.

Now we assume that a static cylindrical black hole is placed in an external asymptotically uniform magnetic field so that the magnetic field lines are parallel to the $ z $ axis. Ricci scalar of the spacetime metric (\ref{metric1}) is proportional to the cosmological constant $R=4 \Lambda$. According to ~\cite{Wald:1974} the nonzero component of the 4-vector potential $ A  {\ mu} $ of the electromagnetic field in space close to the black hole, which is in an external uniform magnetic field, has the following form
\begin{eqnarray}
A^\mu=\frac{1}{2} \left(0, 0, B_0,0 \right)\ \label{potentials1}.
\end{eqnarray}

where $B_0$ is the asymptotic value of the external magnetic field aligned along axis of symmetry, being perpendicular to the equatorial plane where $z =const$. The non zero component of the electromagnetic field tensor (${\cal F}_{\mu\nu}=A_{\nu,\mu}-A_{\mu,\nu}$) is
\begin{eqnarray}\label{FFFF}
{\cal F}_{r \phi}&=&B_0r
\end{eqnarray}

Orthonormal components of the external magnetic field near the BH can be calculated by using the tetrad by proper observer (see for details Eqs.(5)-(9) in Ref.\cite{Rezzolla04}) and non-zero components of the latter have the following form
\begin{equation}\label{BrBt}
    B^{\hat{r}}=B_0 , \\ \qquad B^{\hat{\phi}}=\sqrt{\Delta}B_0\ .
    \end{equation}
    Eq.(\ref{BrBt}) shows that only the angular component of the external magnetic field around the BH reflects gravity effects, while the radial component formally has the same form as in Newtonian case.

\section{charged particles motion}
\begin{figure}
\includegraphics[width=0.45\textwidth]{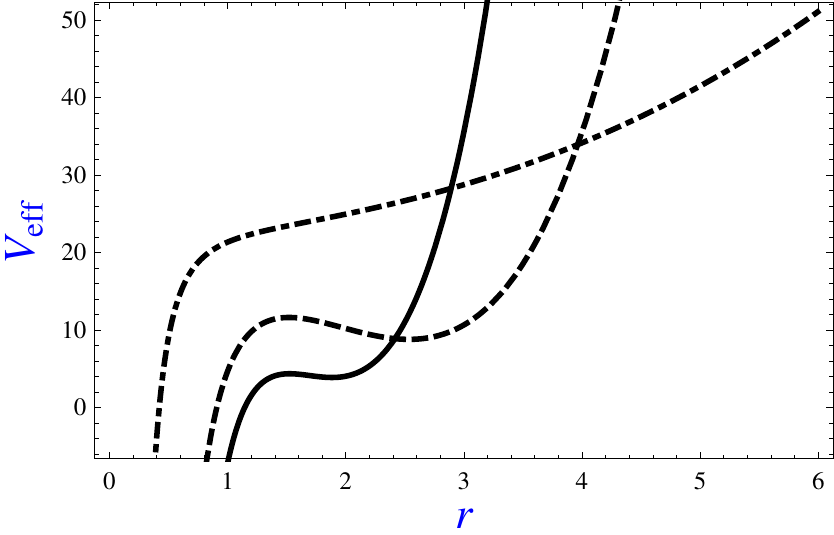}
\caption{\label{effpotfig} The radial dependence of effective
potential for radial motion of charged particles around a cylindrical black hole immersed in an external asymptotically uniform magnetic field for the different values of the dimensionless magnetic interaction parameter: $\omega_B=0.1$ (dot-dashed line), $\omega_B=1$ (dashed line), $\omega_B=5$ (solid line). The mass of central object is taken to be $M=1$, so $\beta=4$ and $\alpha=1.1$. }
  \end{figure}

{Since, the external magnetic field does not break the symmetry of the spacetime,} using the timelike $\xi^\mu_{(t)}$ and spacelike $\xi^\mu_{(\phi)}$  Killing vectors one may easily find two conserved quantities associated with them that are the energy ($E$) and the generalized angular momentum ($L$):

\begin{eqnarray}
\label{Energy}
 E&=&-\xi^\mu_{(t)} (m u_{\mu} + q A_{\mu}) =
m\Delta\frac{dt}{ds}\ ,\\
\label{momentum} L&=&\xi^\mu_{(\phi)} (m u_{\mu} + q A_{\mu}) = m
r^2 \frac {d\phi}{ds} +\frac{qr^2 B_0}{2}\ ,
\end{eqnarray}
where $u^{\mu}$ is the 4-velocity  of the test particle, $s$ is
the affine parameter and $q$ is the charge of the test particle.
By combining these equations with the timelike restriction of test
particles $u_{\mu}u^{\mu}= -1$, one can easily obtain the
equations of motion of the charged particle at equatorial plane
$z=const$ as
\begin{eqnarray}
\label{eom1}
\dot{t}&=&\frac{{\cal E}}{\Delta}\ ,\\
\label{eom11}
\dot{\phi}&=& \frac{{\cal L}}{r^2}-\omega_B\ ,\\
\label{eom2}
\dot{r}^2& =& {\cal E}^2-V_{\rm eff}\ ,
\end{eqnarray}
where the effective potential of radial motion
\begin{equation}
V_{\rm eff}=\Delta\left[1+\left(\frac{{\cal
L}}{r}-\omega_B r \right)^{2}\right]
\end{equation} 
and $\omega_B={qB_0}/{(2mc)}$ is the so-called cyclotron frequency, which is responsible for the magnetic interaction between an electric charge and an external magnetic field,  $ {\cal E}$ and ${\cal L}$ are the energy and momentum per unit mass of the test particle $m$. Fig.~\ref{effpotfig}  the radial dependence of the effective potential of the radial motion of a charged particle in the equatorial plane of a cylindrical black hole immersed in a uniform magnetic field is shown for different values of the magnetic interaction parameter. Now we can draw conclusions about how the parameters of magnetic interaction can change the nature of the motion of charged particles. The parameter of magnetic interaction is responsible for the shift of the minimum value of the effective potential to the center, which means that the minimum distance of charged particles to the center decreases with an increase in the parameter of magnetic interaction. With an increase in the magnetic interaction parameter, parabolic and hyperbolic orbits begin to transform into stable circular orbits. Thus the radial profile of $V_{\rm eff}$ for the different values of the magnetic interaction parameter, running between $0$ and $5$ shows that by increasing the magnetized parameter we also lower the potential barrier, as compared to the pure cylindrical case when the magnetic field is absent.

Next, we will consider the rotational motion of charged particles around a cylindrical black hole, which is influenced by external magnetic fields. The effective potential is of minimum value at the radius of such an orbit. The conditions for the appearance of circular orbits are as follows:
\begin{eqnarray}
\label{condition}\frac{dr}{ds}=0,\quad V_{\rm eff}'=0, \quad
V_{\rm eff}''=0,
\end{eqnarray}
where prime " $'$ " denotes the radial derivative of the functions. Using the first two equations of the [removed]\ ref {condition}) and the [removed]\ ref {eom2}), we can easily derive the following solution to the equation for the energy and angular momentum of a charged particle moving in circular orbits around a black hole located in an external magnetic field:
\begin{eqnarray}\label{energyy}
{\cal E}^2 &=& \left(\alpha ^2 r^2-\frac{\beta}{\alpha  r}\right) \Big\{1+\frac{1}{9 \beta^2}\Big[ 2 \omega_B \left(\alpha ^3 r^4-\beta  r\right)\\\nonumber
&\pm & \sqrt{4 \omega _B^2 \left(\alpha ^3 r^4-\beta  r\right)^2-3 \beta  \left(\beta +2 \alpha ^3 r^3\right)}\Big]^2 \Big \}
\\\label{momentaa}
{\cal L} &=& \frac{r}{3 \beta}\Big\{\omega_B  \left(2 \alpha ^3 r^4+\beta  r\right)
\\\nonumber
&\pm &\sqrt{4 r^2 \omega_B^2 \left(\beta -\alpha ^3 r^3\right)^2-3 \beta  \left(\beta +2 \alpha ^3 r^3\right)}\Big\}\ ,
\end{eqnarray}

\begin{figure}
\includegraphics[width=0.45\textwidth]{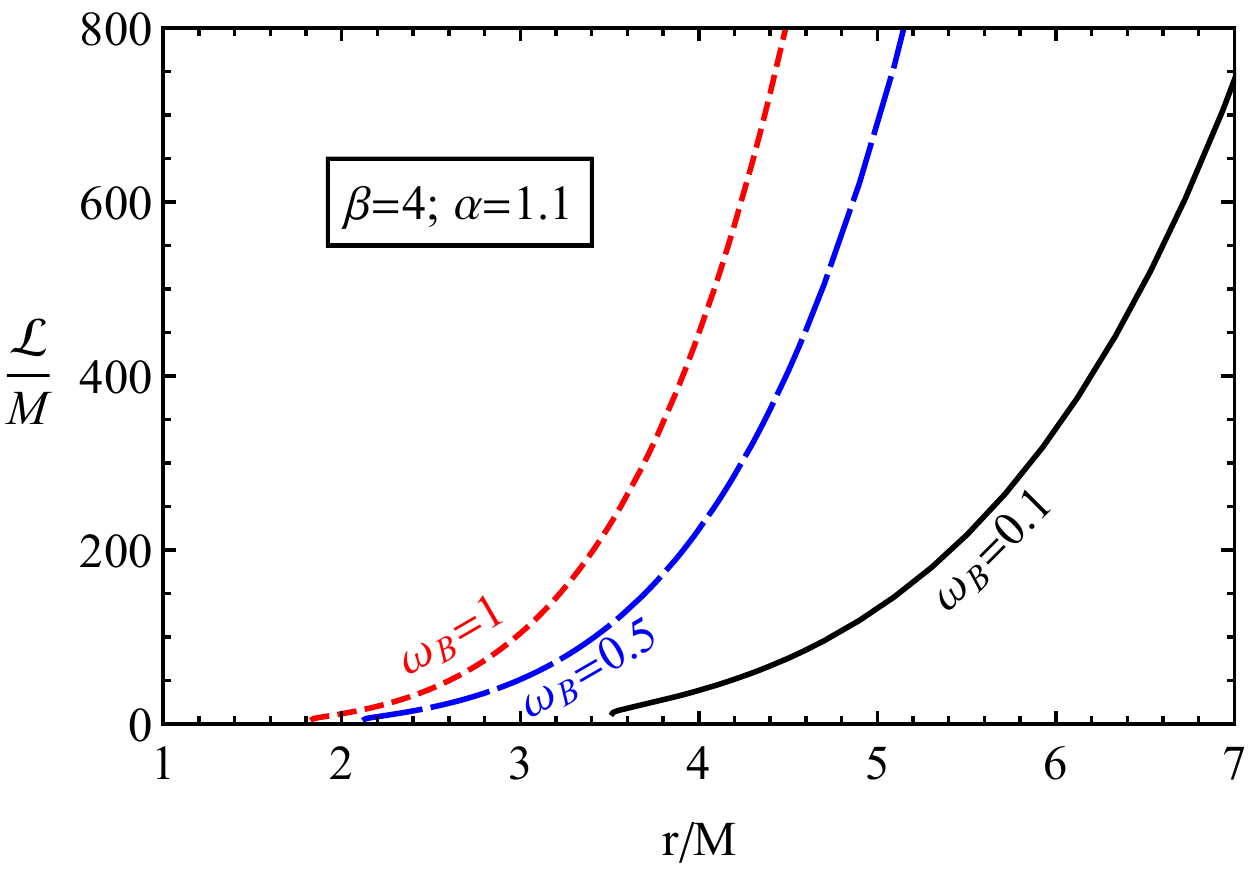}
\includegraphics[width=0.45\textwidth]{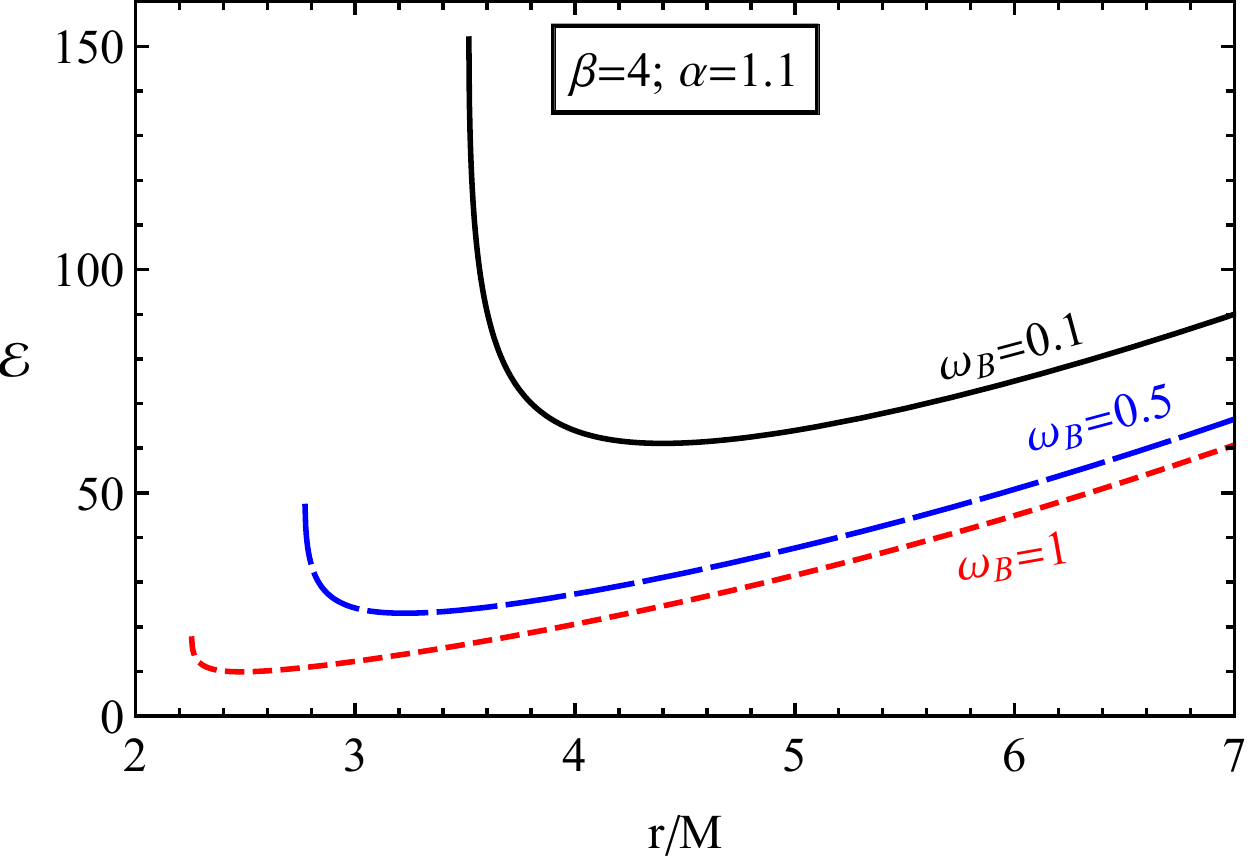}
\caption{\label{angmom} The radial dependence of the angular momentum
(upper) and energy (bottom) of  the charged particle orbiting
around a cylindrical black hole immersed external magnetic field for
the different values of the parameter $\omega_B$: $\omega_B=0.1$ is dashed line,
$\omega_B=1$ is dot-dashed line, $\omega_B=5$ is solid line. The mass of central
object taken to be $M=1$, so $\beta=4$ and $\alpha=1.1$. }
\end{figure}

Fig.~\ref{angmom} shows the radial dependence of both the energy and the angular momenta of the charged particle moving on circular orbits at the equatorial plane. One can easily see that the presence of the magnetic interaction parameter forces test particle to have
bigger energy and angular momentum in order to be kept on the
circular orbit. It is a consequence of the existence of
electromagnetic interaction between uniform magnetic field and
charged test particle in the background gravitational field of the
central object.

In the next stage we will study the ISCO of charged particle
around cylindrical gravitational objects immersed in external
magnetic field. In order to find the values of the ISCO radii
$r_{_{\rm ISCO}}$ one should insert the expressions (\ref{energyy})
and (\ref{momentaa}) to the condition $V_{\rm eff}''=0$ and solve it with respect to the radial coordinate $r$ as

\begin{eqnarray}\nonumber
&&6 \beta  \left(\beta +5 \alpha ^3 r^3\right)-4 r \omega _B \Big[2 \omega _B \\\nonumber && \times  \left(5 \alpha ^6 r^7-7 \alpha ^3 \beta  r^4+2 \beta ^2 r\right)
+\left(\beta +5 \alpha ^3 r^3\right)\\&& \times \sqrt{4 r^2 \omega _B^2 \left(\beta -\alpha ^3 r^3\right)^2-3 \beta  \left(\beta +2 \alpha ^3 r^3\right)}\Big]\geq 0
\end{eqnarray}

Since it is impossible to find analytical solution for
$r_{_{\rm ISCO}}$ we present the results of the numerical solution of
the equation for $r_{_{\rm ISCO}}$ in the Table~\ref{iscotab}. One can
easily see from the results that the presence of the magnetic
field decreases the ISCO radii and forces particle to come closer
to the central object.

\begin{table}
\caption{ \label{iscotab}The dependence of the ISCO radii, energy
and angular momentum of the charged particle from the
dimensionless magnetic interaction parameter $b$.}

       % Give a unique label
\begin{center}
{\begin{tabular}{@{}|c|c|c|c|c|c|c|c|c|@{}} \hline\noalign{\smallskip}
 $\omega_B$ & 0.1 & 0.5 & 1 & 2 & 5 & 10 & 50 \\
\noalign{\smallskip}\hline\noalign{\smallskip}
 ${ r_{_{ISCO}}/M}$ & $6.21$ & $2.95$ & $2.31$ & $1.93$ & $1.65$ & $1.55$ & $1.46$ \\
 \noalign{\smallskip}\hline\noalign{\smallskip}
 ${ \cal E}$ & $13.6$ & $4.16$ &
 $2.78$ & $1.94$ & $1.24$ & $0.88$ & $0.4$   \\
 \noalign{\smallskip}\hline\noalign{\smallskip}
 ${ \cal L}$ & $12.7$ & $4.92$ &
 $4.43$ & $4.99$ & $7.86$ & $12.9$ & $54.6$   \\
 \hline \noalign{\smallskip}
\end{tabular}}
\end{center}
\end{table}
%%%

Now we will analyse the effects of the parameter of cylindrical black hole and magnetic interaction parameters on the ISCO radius of the charged black hole in plot form. 
\begin{figure}
\includegraphics[width=0.45\textwidth]{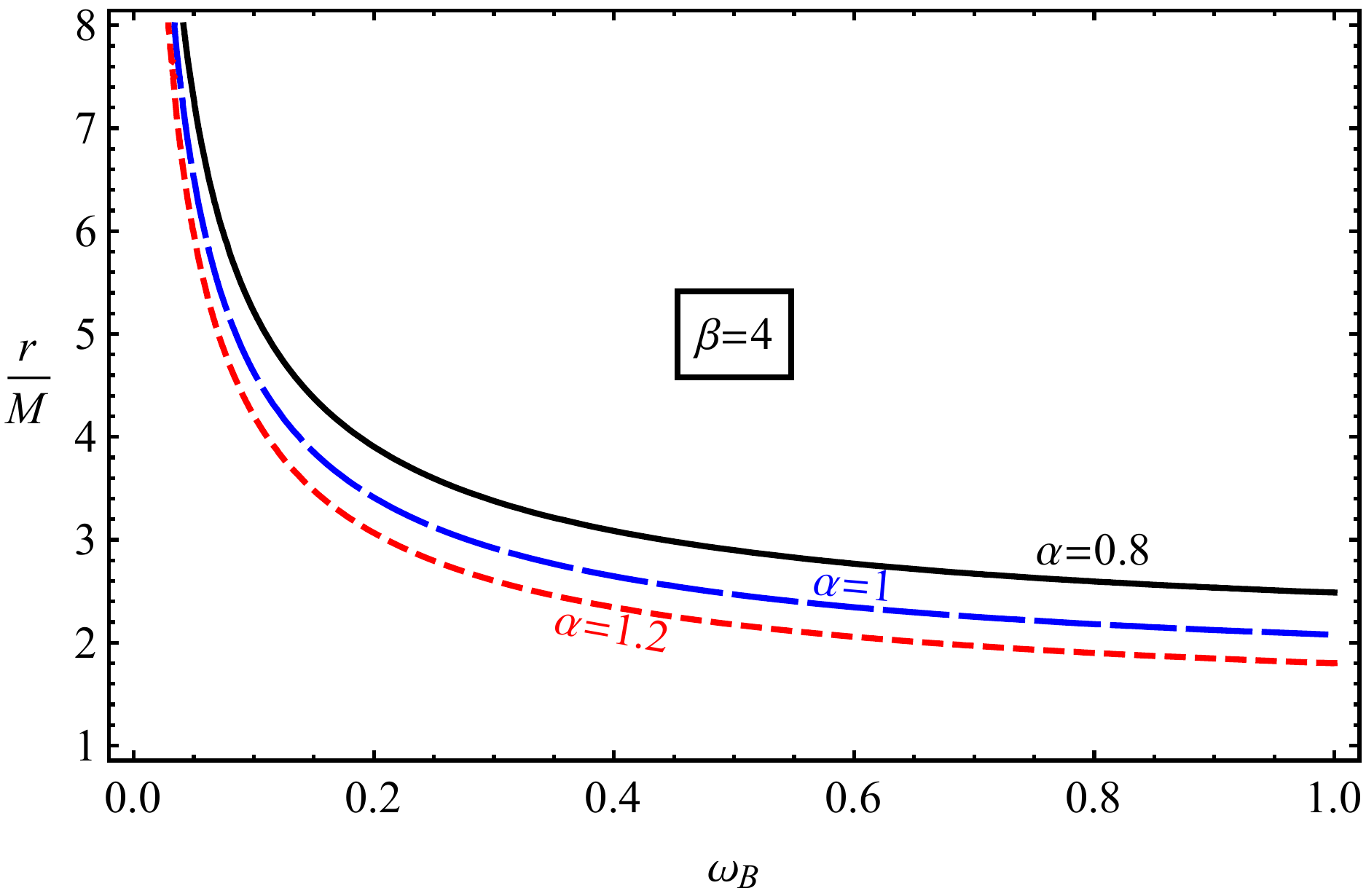}
\caption{\label{iscoch} The dependence of ISCO radius of the charged particle around a cylindrical black hole immersed in the external magnetic fields for the different values of the parameter $\alpha$. }
\end{figure}

Figure ~\ref{iscoch} gives information on the influence of the parameters of a cylindrical black hole and magnetic interaction on the ISCO radius of charged particles rotating around a cylindrical black hole immersed in an external magnetic field. Looking at the figure, we can conclude that an increase in both parameters leads to a decrease in the ISCO radius.
\section {Motion of magnetized particles}
In this part of the study, we focus on the dynamics of magnetized particles around a cylindrical black hole immersed in an external asymptotically uniform magnetic field. The equation of motion of magnetized particles can be expressed using the following Hamilton Jacobi equation \cite{deFelice}
\begin{eqnarray}\label{HJ}
g^{\mu \nu}\frac{\partial {\cal S}}{\partial x^{\mu}} \frac{\partial {\cal S}}{\partial x^{\nu}}=-\Bigg(m-\frac{1}{2} {\cal D}^{\mu \nu}{\cal F}_{\mu \nu}\Bigg)^2\ ,
\end{eqnarray}
where the term ${\cal D}^{\mu \nu}{\cal F}_{\mu \nu}$ stands for the interaction between the magnetized particle and the external magnetic field. Here we assume that the particle has magnetic dipole moment-$\mu^{\nu}$ and the polarization tensor ${\cal D}^{\alpha \beta}$ satisfies the following condition 
\begin{eqnarray}\label{dexp}
{\cal D}^{\alpha \beta}=\eta^{\alpha \beta \sigma \nu}u_{\sigma}\mu_{\nu}\ , \qquad {\cal D}^{\alpha \beta }u_{\beta}=0\ ,
\end{eqnarray}
where $u^{\nu}$ is four velocity of the particle. The electromagnetic field tensor ${\cal F}_{\alpha \beta}$ can be expressed by components of electric $E_{\alpha}$ and magnetic $B^{\alpha}$ fields in the following form
\begin{eqnarray}\label{fexp}
{\cal F}_{\alpha \beta}=2u_{[\alpha}E_{\beta]}-\eta_{\alpha \beta \sigma \gamma}u^{\sigma}B^{\gamma}\ ,
\end{eqnarray}
where square brackets stand for antisymmetric tensor: $T_{[\mu\nu]}=\frac{1}{2}(T_{\mu \nu}-T_{\nu \mu})$ , $\eta_{\alpha \beta \sigma \gamma}$ is the pseudo-tensorial form of the Levi-Civita symbol $\epsilon_{\alpha \beta \sigma \gamma}$ with the relations
\begin{eqnarray}
\eta_{\alpha \beta \sigma \gamma}=\sqrt{-g}\epsilon_{\alpha \beta \sigma \gamma}\, \qquad \eta^{\alpha \beta \sigma \gamma}=-\frac{1}{\sqrt{-g}}\epsilon^{\alpha \beta \sigma \gamma}\ ,
\end{eqnarray}
where $g={\rm det}|g_{\mu \nu}|=-r^4\alpha^2$ for the spacetime metric (\ref{metric}).

The interaction term ${\cal D}^{\mu \nu}{\cal F}_{\mu \nu}$, can be easily defined using the expressions (\ref{dexp}) and (\ref{fexp}) and we have,
\begin{eqnarray}\label{DF1}
{\cal D}^{\mu \nu}{\cal F}_{\mu \nu}=2\mu^{\hat{\alpha}}B_{\hat{\alpha}}=2\mu B_0 \sqrt{\Delta}\ .
\end{eqnarray}

Circular motion of the magnetized particle around the cylindrical black holes is studied by assuming the magnetic moment of the particle is perpendicular to the plane where $z=const$, with the non zero vertical components  $\mu^{i}=(0,0,\mu^{z})$, respectively. The spacetime symmetry does not break the axial symmetric configuration in the presence of the magnetic field and, therefore, one may keep the two conserved quantities: $p_{\phi}= L$ and $p_t = -E$ which are correspond to angular momentum and energy of the magnetized particle, respectively. So, the expression for the action of magnetized particles can be written in the following form

\begin{eqnarray}\label{action}
{\cal S}=-E t+L\phi +{\cal S}_r(r)\ .
\end{eqnarray}
 The form of the action allows to separate variables in the Hamilton-Jacobi equation (\ref{HJ}).

One can easily get the expression of effective potential for radial motion of the magnetized particle at the equatorial plane, with $p_{z}=0$, inserting Eq.(\ref{DF1}) to Eq.(\ref{HJ}) using the form of the action (\ref{action}) and Eq.(\ref{eom2})

\begin{eqnarray}\label{effpot}
V_{\rm eff}(r;\alpha ,l,{\cal B})=\Delta \Big[\left(1-{\cal B} \sqrt{\Delta}\right)^2+\frac{{\cal L}^2}{r^2}\Big]\ ,
\end{eqnarray}
where ${\cal B} = 2\mu B_0/m$ is magnetic coupling parameter being responsible to the magnetic interaction term ${\cal D}^{\mu \nu}{\cal F}_{\mu \nu}$ in the Hamilton-Jacobi equation (\ref{HJ}). In the case when we consider the dynamics of a neutron star with the magnetic dipole moment $\mu=(1/2)B_{\rm NS}R_{\rm NS}^3$, treated as a magnetized particle, orbiting around a supermassive black hole immersed in the external magnetic field, the magnetic coupling parameter $\beta$ can be estimated using the neutron star's observational parameters and the external magnetic field where the neutron star move, around the SMBH
\begin{eqnarray}\label{betaNS}\nonumber
{\cal B} =\frac{B_{\rm NS}R_{\rm NS}^3B_{\rm ext}}{m_{\rm NS}}&\simeq& \frac{\pi}{10^3}\left(\frac{B_{\rm NS}}{10^{12}\rm G}\right)\left(\frac{B_{\rm ext}}{10\rm G}\right)\\
&\times &\left(\frac{R_{\rm NS}}{10^6\rm cm}\right)^3\left(\frac{m_{\rm NS}}{1.4 M_{\odot}}\right)^{-1} .
\end{eqnarray}

\begin{figure}
\includegraphics[width=0.45\textwidth]{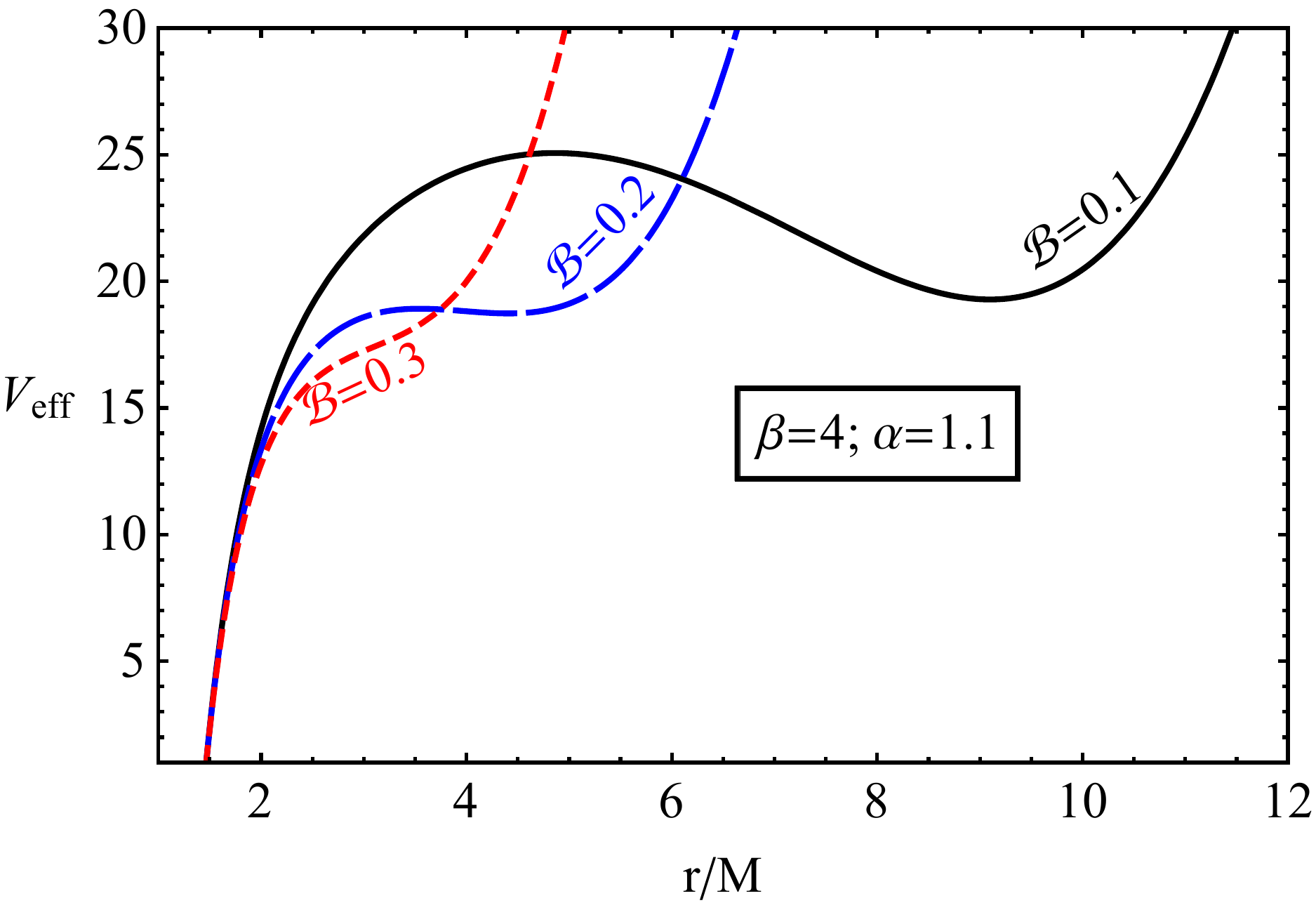}
\caption{\label{effpotmag} The radial dependence of the effective potential of magnetized particles orbiting around the cylindrical black hole immersed in the external magnetic fields for the different values of the parameter ${\cal B}$. The mass of central
object is taken to be $M=1$, so $\beta=4$ and $\alpha=1.1$. }
\end{figure}

Radial dependence of effective potential for radial motion of the magnetized particles for the various values of of the magnetic coupling parameter ${\cal B}$ is presented in Fig.\ref{effpotmag}. One can see from the figure that the maximum and minimum values of the effective potential and the distance from where the potential is maximum decrease with the increase of the magnetic coupling parameter.

One can define circular orbits of particles around a black hole by the following conditions
\begin{eqnarray} \label{conditions}
\dot{r}=0 \ , \qquad \frac{\partial V_{\rm eff}}{\partial r}=0 \ .
\end{eqnarray}
From the conditions (\ref{conditions}) 

\begin{eqnarray}
&&{\cal L}=\frac{r^3 \Delta '}{2 \Delta -r \Delta'}\left(1-3 {\cal B} \sqrt{\Delta}+2 {\cal B}^2 \Delta \right) \ , \\
&&{\cal E}=\frac{\Delta^{\frac{3}{2}}\left(1-{\cal B}  \sqrt{\Delta}\right)}{2 \Delta -r \Delta'} \left[2 \sqrt{\Delta}-{\cal B} ( r \Delta' +2 \Delta)\right] \ .
\end{eqnarray}

\begin{figure}
\includegraphics[width=0.45\textwidth]{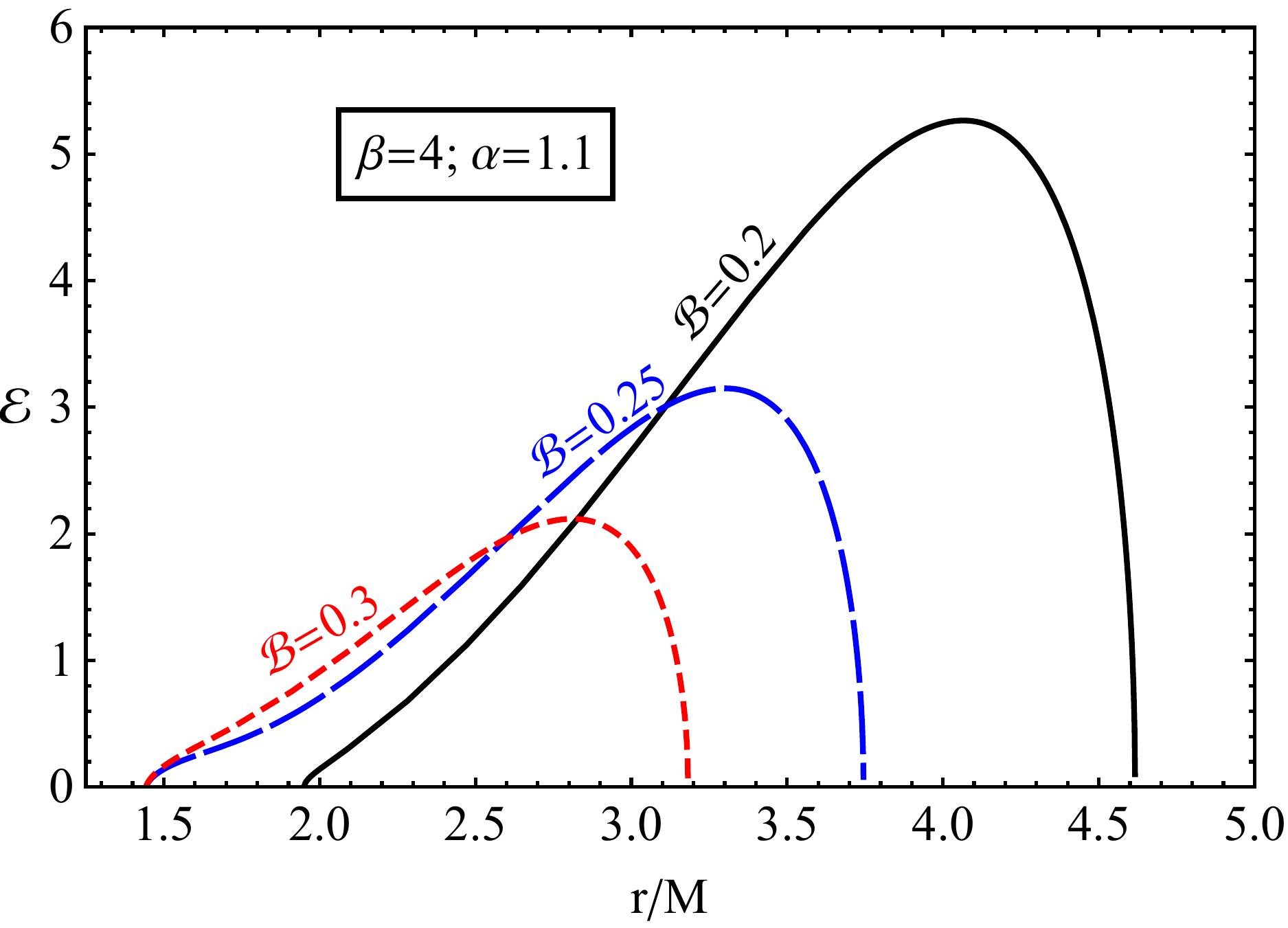}
\includegraphics[width=0.45\textwidth]{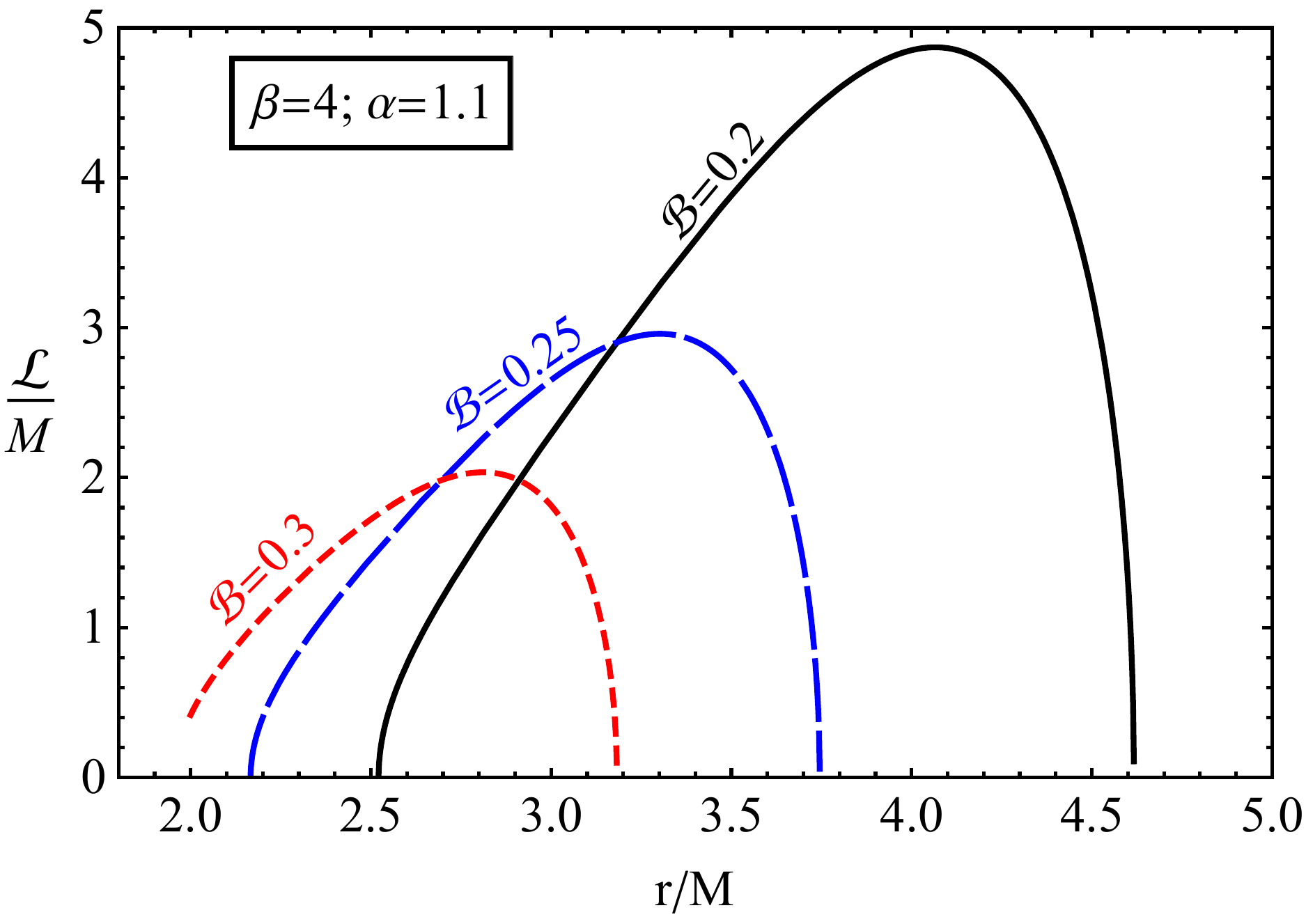}
\caption{\label{angenmag} The radial dependences of the energy (top graph) and angular momentum (bottom graph) of magnetized particles rotating in circular orbits around a black hole in an external magnetic field, taken at different values of the parameters of the relations $ {\ cal B} $.}
\end{figure}

Figure ~\ref{angenmag} demonstrates the radial dependence of the specific energy and angular momentum for the circular motion of magnetized particles around a cylindrical black hole placed in an external magnetic field, taken at different values of the magnetic coupling parameter. An increase in the magnetic coupling parameter leads to a decrease in both energy and angular momentum, which can be seen from the figures.

ISCO equation finds using the standard condition $V_{rr}\geq0$ and we have got the following equation which its solution with respect to the radial coordinate to be ISCO radius

ISCO equation finds using the standard condition $V_{rr}\geq0$ and we have got the following equation which its solution with respect to the radial coordinate to be ISCO radius

\begin{eqnarray}\label{iscoeqmag}
\nonumber
 {\cal B} r^2 \left( 4 {\cal B} \sqrt{\Delta}-3 \right) \Delta'{}^2
+2 r \sqrt{\Delta} \left(2 \Delta -3 {\cal B}  \sqrt{\Delta}+1 \right)\\-4 r \Delta^{3/2} \left(2 {\cal B}^2 \Delta -3 {\cal B}  \sqrt{\Delta}+1 \right)\frac{\Delta ''}{\Delta '} \geq 0
\end{eqnarray}
 
 From the equation (\ref{iscoeqmag}) it can be seen that it is impossible to solve it analytically with respect to the radial coordinate and to see the effects of the interaction of the magnetic dipole of magnetized particles and the external magnetic field, as well as the parameter $\alpha$. One way to analyze this numerically is to plot ISCO as a function of the parameter $\alpha$.

\section{Acceleration of particles near cylindrical black holes}

Observations show that the brightness of galactic center where a supermassive black holes is situated, much more brighter than other places of the galaxy. By the reason, the study of energy release processes from different black holes with different ways is always one of interesting issues in relativistic astrophysics. For the first time Rodger Penrose has suggested a new mechanism of the energetic process from rotating Kerr black hole which has ergoregion where a particle come and decays to two, one of the parts falls down to the central black hole with negative energy and other one goes to infinity having bigger energy than initial one~\cite{Penrose:1963} . The Penrose process have been developed during the past years by several authors applying rotating black holes in different gravity theories and conditions~\cite{Dadhich18,wagh85}. Another model for the energetic process is proposed by Banados, Silk and West produced by collisions of particles near horizon of the black hole \cite{Banados09,Abdujabbarov13a}. Now, here we investigate the collisions of neutral, charged and magnetized particles and find the energy $E_{\rm cm}$ in the center of mass of
the system with energy  at infinity $E_{1}$ and $E_{2}$ in the gravitational field described by spacetime metric (\ref{metric1}) in the presence of external
asymptotically uniform magnetic field. We will use the following 
expression~\cite{Banados09}
\begin{equation}
\left(\frac{1}{\sqrt{-g_{00}}} \ E_{\rm cm}, 0, 0, 0\right)=
m_{1}u_{(1)}^{\mu}+m_{2}u_{(2)}^{\nu},
\end{equation}
where $u_{(1)}^{\mu}$ and $u_{(2)}^{\nu}$ are the 4-velocities of
the particles, properly normalized by $g_{\mu\nu}u^{\mu}u^{\nu}=-1$ and $m_{1}$, $m_{2}$ are the rest masses of the particles, to find the expression for the center of mass energy. We will consider two particles with equal mass ($m_{1}=m_{2}=m$)  which have the energy at infinity ${\cal E}_{1}={\cal E}_{2} = 1$. Thus we have
\begin{eqnarray}\label{ECM}
{\cal E}_{cm}=\frac{E_{\rm cm}}{\sqrt{2}m}=  \sqrt{1- g_{\mu\nu} u_{(1)}^{\mu}
u_{(2)}^{\nu}}.
\end{eqnarray}

In below we will study effects of spacetime around cylindrical black holes on energetics of collisions of neutral, charged and magnetized particles in frame of proper observer.

\subsection{Neutral particles collisions}

The expression for center of mass energy of the two neutral particles can be calculated by substituting Eqs.(\ref{eqmotionm}) for the case ${\cal B}=0$ in to Eq.(\ref{ECM}) in the following form:
\begin{eqnarray}\label{ecmnneq}
\nonumber
{\cal E}_{\rm cm}^2&=&1+\frac{1}{\Delta}-\frac{l_1l_2}{r^2}-\sqrt{1-\Delta \left(1+\frac{l_1^2}{r^2}\right)}
\\
&\times &\frac{1}{\Delta} \sqrt{1-\Delta \left(1+\frac{l_2^2}{r^2}\right)}\ .
\end{eqnarray}
Now, we will analyse effects of the parameter $\alpha$ on the center of mass energy of collisions of two neutral particles around cylindrical black hole plotting Eq.(\ref{ecmnneq}).
%%%%%%%%%%%%%%%
\begin{figure}[ht!]
\centering
\includegraphics[width=0.98\linewidth]{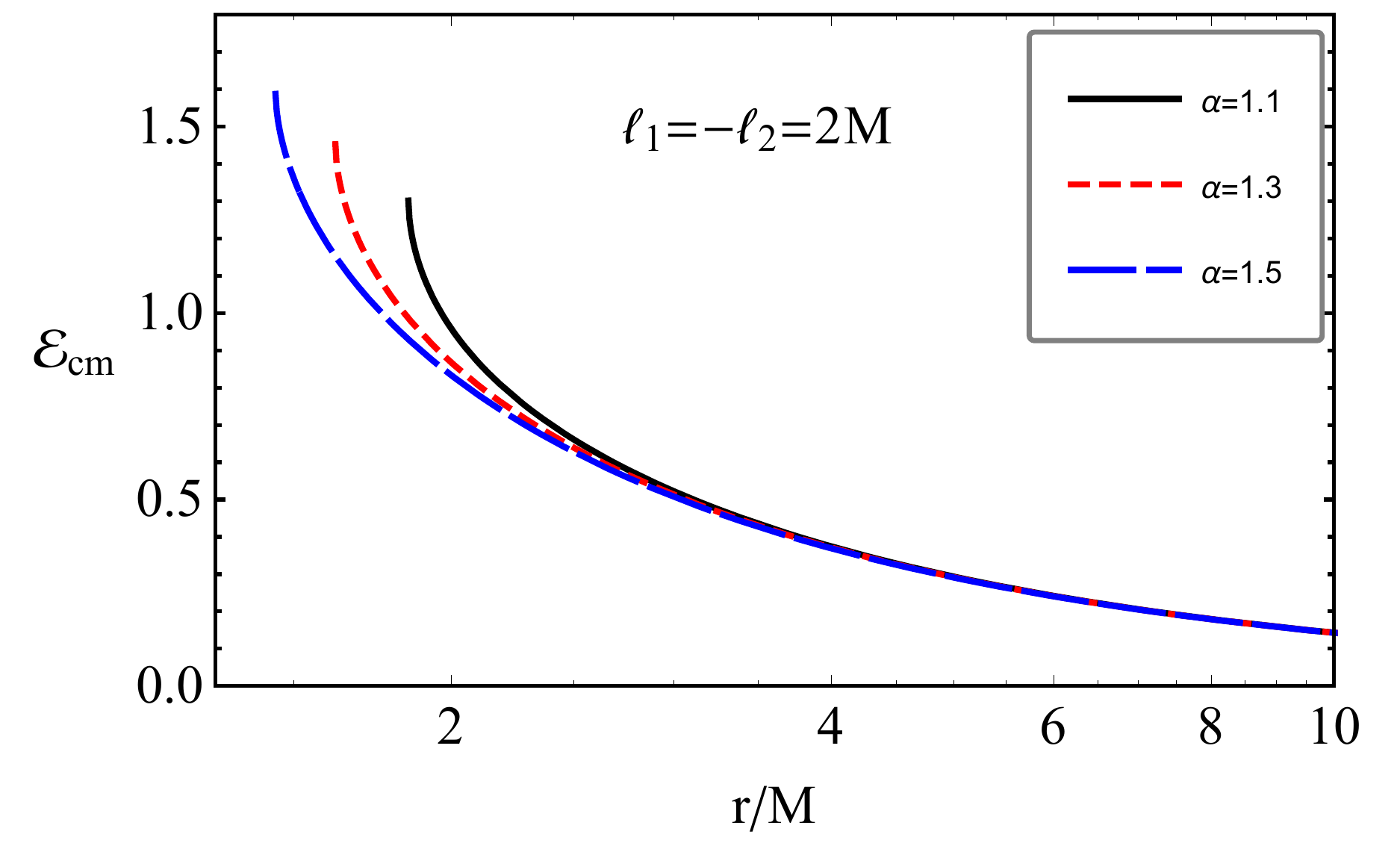}
\caption{Radial dependence of the center of mass energy of collision of two neutral particles around cylindrical black hole for the different values of the parameter $\alpha$.  \label{centernn}}
\end{figure}

Figure ~\ref{centernn} shows the radial dependence of the energy of the center of mass of neutral particles rotating around a static black hole with cylindrical symmetry for various values of the parameter $\alpha$ for the values of the specific energy of colliding particles $l_1=2M$ and $l_2=2M$. It is seen that the energy of the center of mass increases due to the growth of the parameter $\alpha$.

\subsection{Charged particles collisions}

Now consider the collisions of two charged particles in the vicinity of a cylindrical black hole immersed in an external magnetic field. Applying equations (\ref{eom1})-(\ref{eom2}), we can obtain equations for the energy of the center of mass of colliding particles in the following form:
\begin{eqnarray} \label{CME1}
\nonumber
{\cal E}_{\rm cm}^2&=&
1+\frac{1}{\Delta}-
\left(\frac{\ell_{1}}{r}-\omega_B^{(1)}\right)
\left(\frac{\ell_{2}}{r}-\omega_B^{(2)}\right)\\ \nonumber &-& \frac{1}{\Delta} \sqrt{1-\Delta \left[1+ \left(\frac{\ell_{1}}{r}- \omega_B^{(1)}r \right)^{2}\right]} \\&& \times \sqrt{1-\Delta \left[1+ \left(\frac{\ell_{2}}{r}- \omega_B^{(2)}r \right)^{2}\right]}.
\end{eqnarray}
Now we need to analyze the influence of space-time around the black hole and the magnetic field on the energy of the center of mass of collisions of positive-positive, negative-negative and positive-negative charged particles by writing the equation (\ref{CME1}).
\begin{figure}
\includegraphics[width=0.45\textwidth]{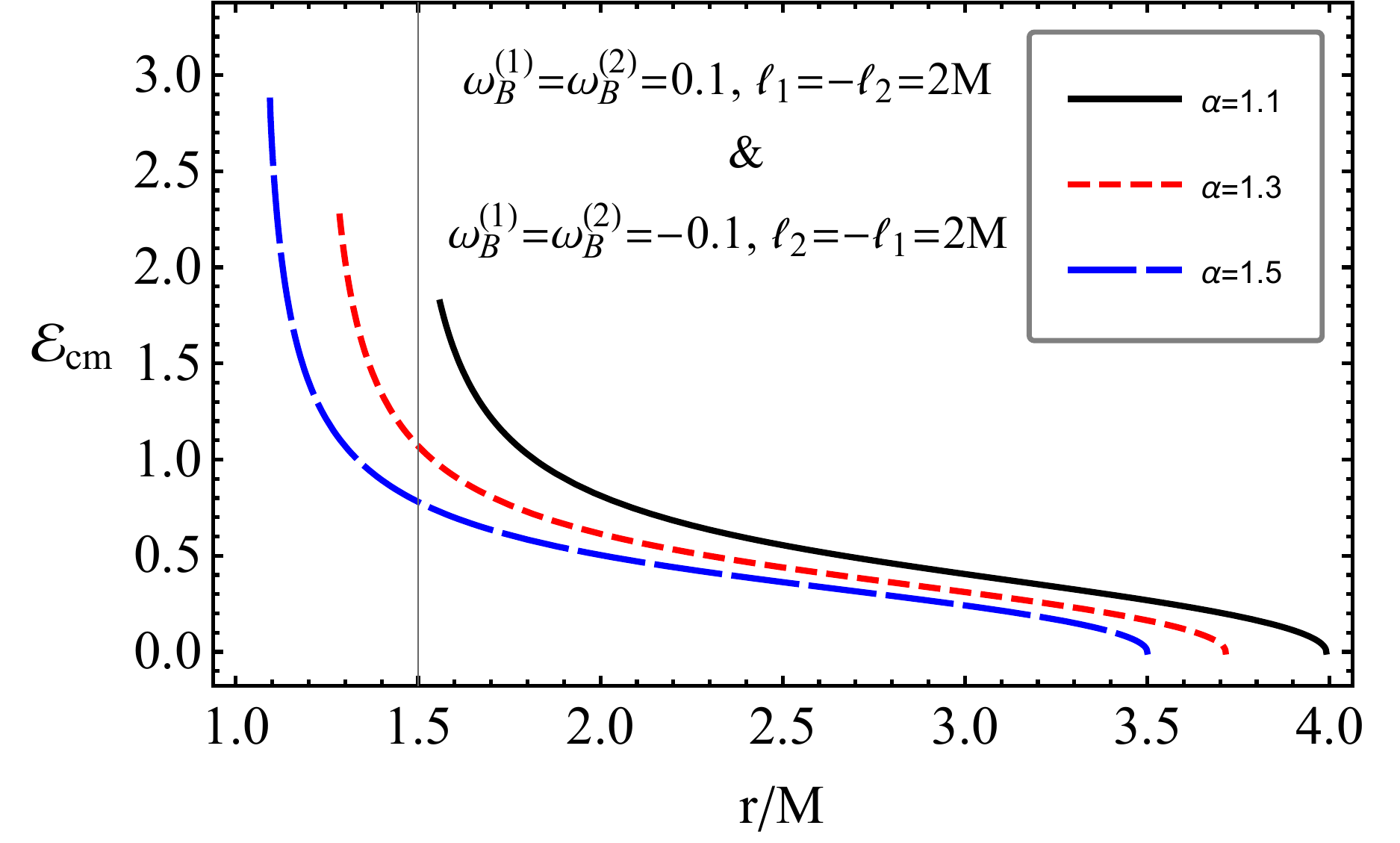}
\includegraphics[width=0.45\textwidth]{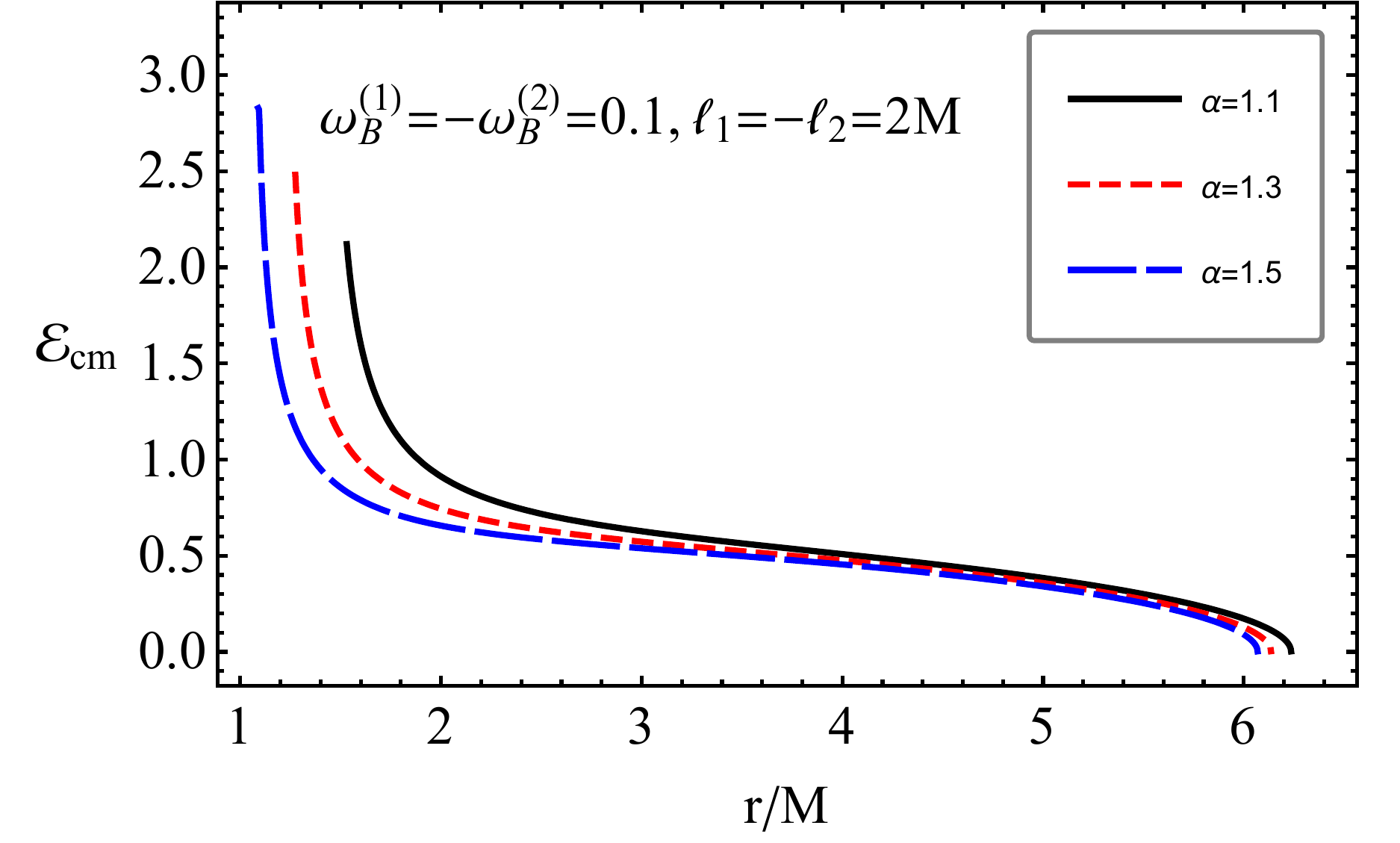}
\caption{\label{ecm}The radial dependence of the center of mass on the collision energy of two charged particles around a black hole placed in an external asymptotically uniform magnetic field, which are taken for different values of the parameter $\alpha$.}
\end{figure}

In fig.\ref{ecm} shows the radial dependence of the center-of-mass energy of two particles taken at different values of the dimensionless parameter ${\cal B}$. It can be seen from the figure that in the presence of a magnetic field, a particle can significantly increase the acceleration process near the horizon to high energies. In addition, we show that the radial dependence of the center of mass energy in the collision of positive-positive and negative-negative charged particles is the same for the case when the orientation of the particles is reversed.

\subsection{Magnetized particles collisions}
In this subsection we consider collisions of two magnetized particles.
The four-velocity of the magnetized particle at equatorial plane ($z=const,\ \dot{z}=0$) has the following components:
\begin{align}\label{eqmotionm}
\nonumber
\dot{t}&=\frac{{\cal E}}{\Delta}\ ,
\\
\nonumber
\dot{r}^2&={\cal E}^2-\Delta \left[ \left(1-{\cal B} \sqrt{\Delta} \right)^2+\frac{l^2}{r^2}\right] \ ,
\\
\dot{\phi}&=\frac{l}{r^2}\ .
\end{align}
One could find the expression for center of mass energy of the two magnetized particles inserting Eq.(\ref{eqmotionm}) in to Eq.(\ref{ECM}) in the following form:
\begin{eqnarray}
\nonumber
{\cal E}_{\rm cm}^2&=&1+\frac{1}{\Delta}-\frac{l_1l_2}{r^2}-\sqrt{1-\Delta \left[ \left(1-{\cal B}_1\sqrt{\Delta} \right)^2+\frac{l_1^2}{r^2}\right]}
\\
&\times &\frac{1}{\Delta} \sqrt{1-\Delta \left[\left(1-{\cal B}_2\sqrt{\Delta}\right)^2+\frac{l_2^2}{r^2}\right]}\ .
\end{eqnarray}
%%

%%%%%%%%%%%%%%%
\begin{figure}[ht!]
\centering
\includegraphics[width=0.98\linewidth]{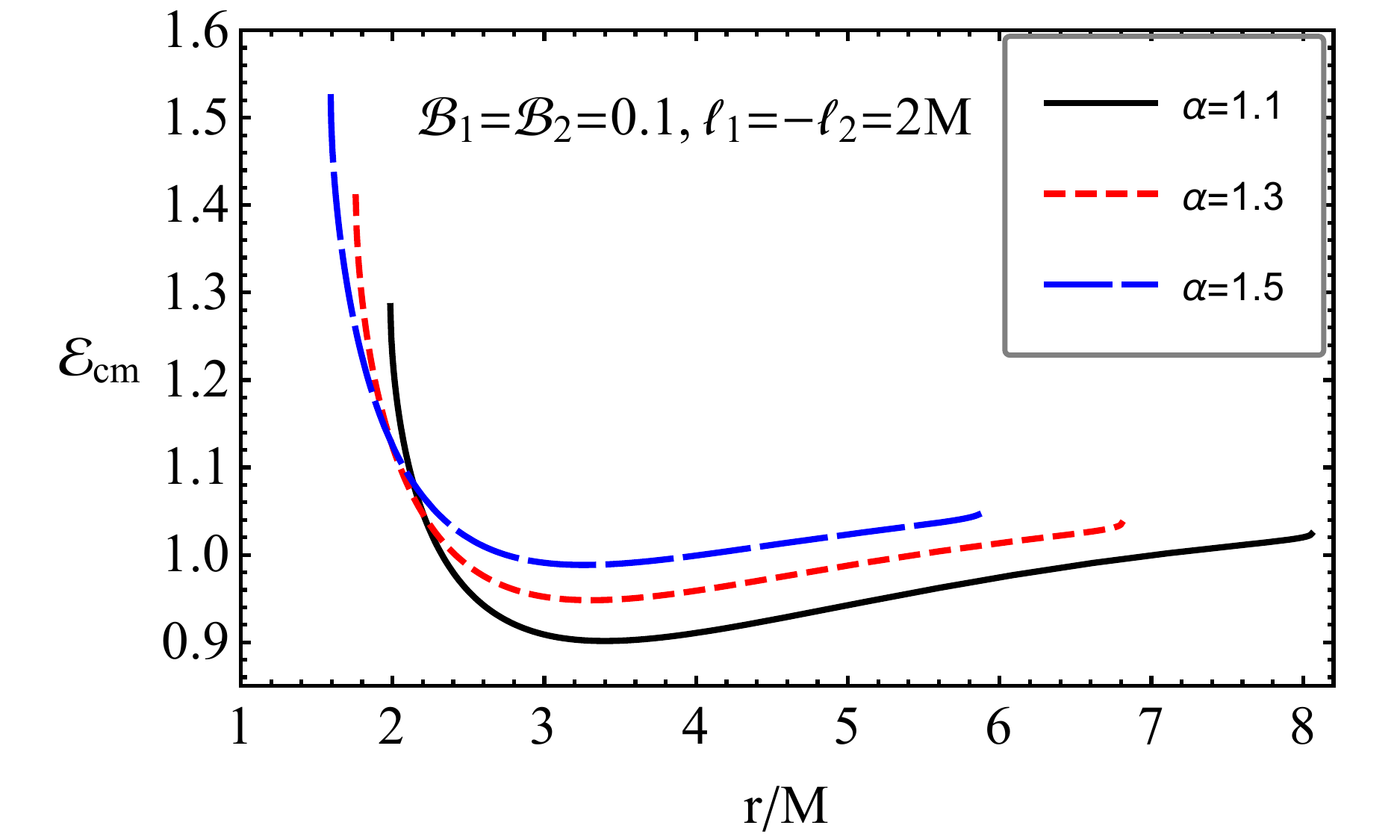}
\includegraphics[width=0.98\linewidth]{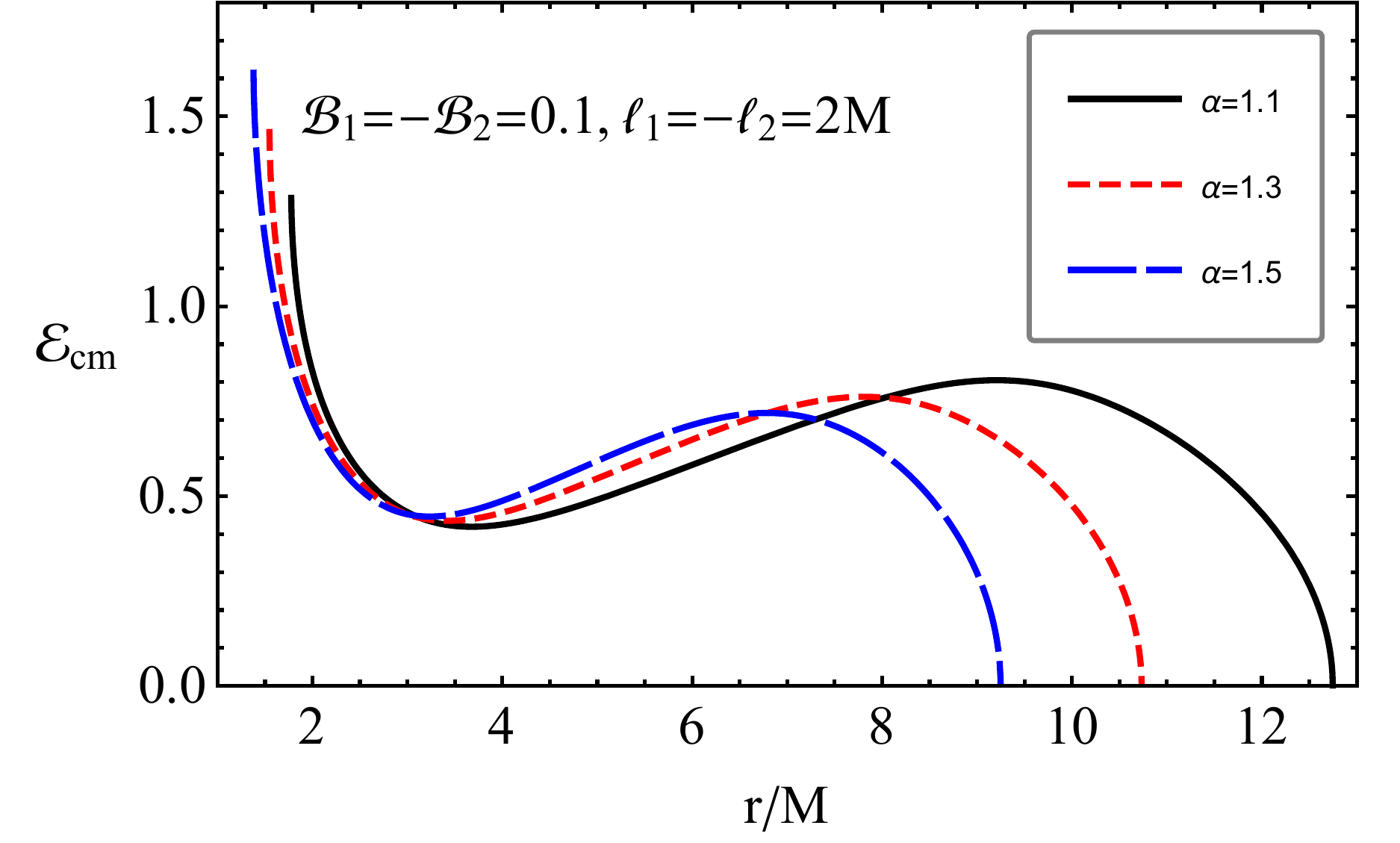}
\includegraphics[width=0.98\linewidth]{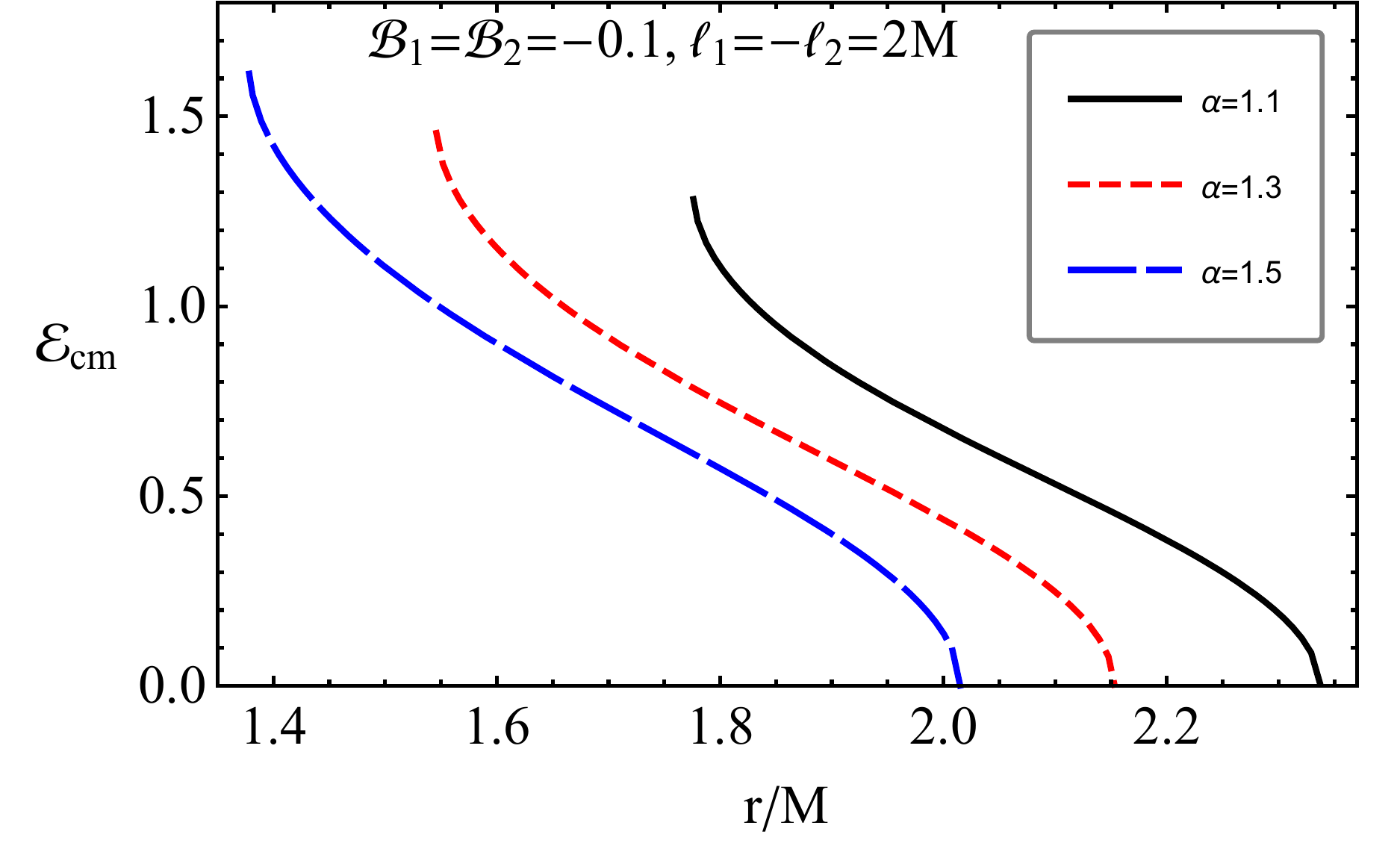}
\caption{Radial dependence of the center of mass energy of collision of two magnetized particles with the same initial energy ${\cal E}_1={\cal E}_2=1$, around 4-D EGB BH for the different values of the $\alpha$ parameter.  \label{centermm}}
\end{figure}

Figure ~\ref{centermm} shows information on the radial dependence of the center of mass energy of magnetized particles rotating around a cylindrical black hole in an asymptotically uniform external magnetic field at different $\alpha$. To plot Fig.\ref{centermm} values of the specific angular momentum for colliding magnetized particles are chosen as $ l_1 / M = 2 $ and $l_2/M =-2$, and also collisions of magnetized particles with co-directional and oppositely directed magnetic dipoles, having values of positive and negative parameters of magnetic interaction. The upper graph gives information that the collision energy of the center of mass of magnetized particles with the same direction of magnetic dipoles and external magnetic field increases with increasing value of the parameter $ \ alpha $. It should be noted that at a large distance the center of mass of energy disappears, this means that the collision of particles does not occur due to the predominant effect of magnetic interaction between the dipoles of the particles and the minimum distance at which the particles cannot collide with each other and the center of mass energy disappearance comes close to the central object with the increase of the parameter $\alpha $ The middle panel of the figure demonstrates the case of collision of magnetized particles with the opposite direction of magnetic dipoles, when the energy of the center of mass has a minimum and a maximum, in other words, with distance, the energy decreases due to to reduce the gravitational potential, and then it increases again when the magnetic interaction between an external magnetic field and magnetic dipoles plays an important role, and finally it decreases and tends to zero due to the dominant interaction effect between the two dipoles. Finally, in the case when the magnetic dipole moment of both the colliding magnetized particles and the directions of the external magnetic field are opposite, the energy of the center of mass decreases faster than in the cases considered above due to the repulsive properties of the magnetic interaction between the particles and the external magnetic field.
\section{conclusion}
\label{conclusions}

In this article, we studied the motion of a charged particle around a black hole in cylindrical coordinates in the presence of an external asymptotically uniform magnetic field. The effective potential of the radial motion of a test particle in the equatorial plane is studied for the case of various values  of the magnetized parameter responsible for the interaction of the magnetic field and charged particles in the background gravitational field. Using the conditions for the minimum form of the effective potential, we obtained the numerical values  of the energy, angular momentum, and radii of ISCO. We have shown that the presence of a magnetic field can reduce the radius of the ISCO, and charged particles can move closer to the center of the black hole. Decreasing the ISCO radius is very important because the gravitational potential near the central object can accelerate particles to high energies.

In Fig.~\ref{effpotfig}  we have shown the radial dependence of the
effective potential of the radial motion of a charged particle at an equatorial plane of a cylindrical black hole immersed in an external asymptotically uniform magnetic field for different values of the magnetic interaction parameter $\omega_B$. We conclude that the  minimum distance of the charged particles to the central object decreases with the increase of the magnetic interaction parameter. With the increase of the magnetic interaction parameter parabolic and hyperbolic orbits start to become stable circular orbits.

Then we have shown in Fig.~\ref{angmom} the radial dependence of
both the energy and the angular momenta of the charged particle
moving on circular orbits in the equatorial plane. The presence of
the magnetic interaction forces charged test particle to have bigger
energy and angular momentum in order to be kept on its circular
orbit, which is the consequence of the existence of magnetic interaction between the external magnetic field and the test charged particle in the background gravitational field of the
central object.

We also investigate dynamics of magnetized particles around the cylindrical black hole immersed in the external magnetic fields and shown that the maximum values of the specific energy and angular momentum decreases with the increase of the magnetic coupling parameter. ISCO radius decrease with increase of the parameter $\alpha$ for the magnetized particle with the magnetic coupling parameter ${\cal B}=0.1$.

Authors of~\cite{Banados09} underlined that a rotating black
hole can, in principle, accelerate the particles falling to the
central black hole to arbitrary high energies. The open question
is whether there is additional effects beside the rotation which could
play a role of particle accelerator near the black hole. Here we
have investigated neutral, electrically charged and magnetized particle acceleration mechanism near the non-rotating cylindrical black hole in the presence of the external magnetic field. We have obtained the exact analytical expression for center of mass energy  $E_{\rm cm} $ of the two particles. In the
Fig.~\ref{ecm} the radial dependence of the $E_{\rm cm}$ has been
shown for the different values of the dimensionless magnetic interaction  parameter $\omega_B$. One can see from this figure that  in the presence of the external magnetic field the particles could be accelerated to much higher energies with compare to the case when the magnetic field is absent and we shown that in cases when the two positive-positive and negative-negative charged particles collisions the behaviour of the radial dependence of the center of mass energy has symmetry in the following replacements:$\left(\omega_B^{(1)},\omega_B^{(2)}\right) \to \left(-\omega_B^{(1)},-\omega_B^{(2)}\right)$ and $l_1 \to -l_1; -l_2 \to l_2 $.

\begin{acknowledgements}

This research is supported by Grants No. VA-FA-F-2-
008 and No. MRB-AN-2019-29 of the Uzbekistan Ministry
for Innovative Development. 

\end{acknowledgements}

\bibliographystyle{apsrev4-1}
\bibliography{gravreferences}

%
%\subsection{astrophysical application to the SMBH - SGR A*}

\end{document}